\documentclass[journal]{IEEEtran}

\usepackage{bm,amsmath,amssymb,mathrsfs,amsfonts}
\usepackage{caption}
\usepackage{scalefnt,multirow}
\usepackage{here}
\usepackage{booktabs,arydshln}
\usepackage{enumitem}
\usepackage{type1cm}
\usepackage{cite}
\usepackage{url}
\usepackage{algorithm,algpseudocode}
\usepackage{color}
\usepackage{graphicx}
\usepackage{tikz}
\usepackage{pgfplots}
\pgfplotsset{compat=1.17}

\newcommand{\ie}{\textit{i.e.}}

\newcommand{\etal}{\textit{et~al.}}

\newcommand{\proposed}{Orthros}
\newcommand{\ar}{autoregressive}

\newcommand{\argmax}{\mathop{\rm argmax}\limits}

\newcommand{\dmodel}{d_{\rm model}}
\newcommand{\dff}{d_{\rm ff}}
\newcommand{\nhead}{H}
\newcommand{\speech}{X}
\newcommand{\xmax}{U}
\newcommand{\ysrc}{Z}
\newcommand{\ytgt}{Y}
\newcommand{\ymax}{N}
\newcommand{\ymaxhat}{\hat{N}}
\newcommand{\vocab}{{\cal V}}
\newcommand{\masktoken}{\texttt{[MASK]}}
\newcommand{\lengthtoken}{\texttt{[LENGTH]}}
\newcommand{\yobs}{Y_{\rm obs}}
\newcommand{\yobsmmt}{Y_{\rm obs}^{(m)}}
\newcommand{\ymask}{Y_{\rm mask}}
\newcommand{\ymaskmmt}{Y_{\rm mask}^{(m)}}
\newcommand{\nmask}{M}
\newcommand{\yhat}{\hat{Y}}
\newcommand{\yhatobs}{\hat{Y}_{\rm obs}}

\newcommand{\itert}{(t)}
\newcommand{\itertminusone}{(t-1)}
\newcommand{\totaliter}{T}
\newcommand{\ndecar}{N_{\rm \textsc{ar}}}
\newcommand{\beamst}{b_{\rm \textsc{st}}}

\newcommand{\lengthbeam}{l}
\newcommand{\lambdactc}{\lambda_{\rm \textsc{ctc}}}
\newcommand{\lambdalp}{\lambda_{\rm lp}}
\newcommand{\lambdaar}{\lambda_{\rm \textsc{ar}}}
\newcommand{\lambdamt}{\lambda_{\rm \textsc{mt}}}
\newcommand{\losstotal}{{\cal L}_{\rm total}}
\newcommand{\lossar}{{\cal L}_{\rm \textsc{ar}}}
\newcommand{\lossctc}{{\cal L}_{\rm \textsc{ctc}}}
\newcommand{\losscmlm}{{\cal L}_{\rm \textsc{cmlm}}}
\newcommand{\losslp}{{\cal L}_{\rm lp}}
\newcommand{\lossmt}{{\cal L}_{\rm \textsc{mt}}}
\newcommand{\probar}{P_{\rm \textsc{ar}}}
\newcommand{\probnar}{P_{\rm \textsc{nar}}}
\newcommand{\probctc}{P_{\rm \textsc{ctc}}}
\newcommand{\problp}{P_{\rm lp}}
\newcommand{\probcmlm}{P_{\rm \textsc{cmlm}}}

\usepackage{pifont}%

\usepackage{etoolbox}

\newcommand{\algstrut}[1][\algruledefaultfactor]{\vrule width 0pt
depth .25\baselineskip height #1\baselineskip\relax}
\newcommand*{\algrule}[1][\algorithmicindent]{\hspace*{.5em}\vrule\algstrut
\hspace*{\dimexpr#1-.5em}}

\makeatletter
\newcount\ALG@printindent@tempcnta
\def\ALG@printindent{%
    \ifnum \theALG@nested>0%
    \ifx\ALG@text\ALG@x@notext%
    \else
    \unskip
    \ALG@printindent@tempcnta=1
    \loop
    \algrule[\csname ALG@ind@\the\ALG@printindent@tempcnta\endcsname]%
    \advance \ALG@printindent@tempcnta 1
    \ifnum \ALG@printindent@tempcnta<\numexpr\theALG@nested+1\relax%
    \repeat
    \fi
    \fi
}%

\patchcmd{\ALG@doentity}{\noindent\hskip\ALG@tlm}{\ALG@printindent}{}{\errmessage{failed to patch}}

\AtBeginEnvironment{algorithmic}{\lineskip0pt}

\makeatletter
\def\adl@drawiv#1#2#3{%
        \hskip.5\tabcolsep
        \xleaders#3{#2.5\@tempdimb #1{1}#2.5\@tempdimb}%
                #2\z@ plus1fil minus1fil\relax
        \hskip.5\tabcolsep}
\newcommand{\cdashlinelr}[1]{%
  \noalign{\vskip\aboverulesep
           \global\let\@dashdrawstore\adl@draw
           \global\let\adl@draw\adl@drawiv}
  \cdashline{#1}
  \noalign{\global\let\adl@draw\@dashdrawstore
           \vskip\belowrulesep}}
\makeatother

\begin{document}
\title{Non-autoregressive End-to-end Speech Translation with Parallel Autoregressive Rescoring}

\author{Hirofumi~Inaguma,~\IEEEmembership{Student Member,~IEEE,}
        Yosuke~Higuchi,~\IEEEmembership{Student Member,~IEEE,}\\
        Kevin~Duh,~\IEEEmembership{Member,~IEEE}
        Tatsuya~Kawahara,~\IEEEmembership{Fellow,~IEEE}
        and~Shinji~Watanabe,~\IEEEmembership{Senior Member,~IEEE}%
\thanks{H. Inaguma and T. Kawahara are with the
Graduate School of Informatics, Kyoto University, Kyoto 606-8501, Japan (e-mail: \{inaguma, kawahara\}@sap.ist.i.kyoto-u.ac.jp). \par
Y. Higuchi is with Waseda University, Tokyo 162-0042, Japan (e-mail: higuchi@pcl.cs.waseda.ac.jp). \par
K. Duh is with HLTCOE, Johns Hopkins University, Baltimore, MD 21211-2840, USA (e-mail: kevinduh@cs.jhu.edu). \par
S. Watanabe is with Language Technologies Institute, Carnegie Mellon University, Pittsburgh, PA 15213-3891, USA (e-mail: shinjiw@cmu.edu).
}%
}

\maketitle

\begin{abstract}
This article describes an efficient end-to-end speech translation (E2E-ST) framework based on non-autoregressive (NAR) models.
End-to-end speech translation models have several advantages over traditional cascade systems such as inference latency reduction.
However, conventional AR decoding methods are not fast enough because each token is generated incrementally.
NAR models, however, can accelerate the decoding speed by generating multiple tokens in parallel on the basis of the token-wise conditional independence assumption.
We propose a unified NAR E2E-ST framework called \textit{Orthros}, which has an NAR decoder and an auxiliary shallow AR decoder on top of the shared encoder.
The auxiliary shallow AR decoder selects the best hypothesis by rescoring multiple candidates generated from the NAR decoder in parallel (\textit{parallel AR rescoring}).
We adopt conditional masked language model (CMLM) and a connectionist temporal classification (CTC)-based model as NAR decoders for Orthros, referred to as Orthros-CMLM and Orthros-CTC, respectively.
We also propose two training methods to enhance the CMLM decoder.
Experimental evaluations on three benchmark datasets with six language directions demonstrated that Orthros achieved large improvements in translation quality with a very small overhead compared with the baseline NAR model.
Moreover, the Conformer encoder architecture enabled large quality improvements, especially for CTC-based models.
Orthros-CTC with the Conformer encoder increased decoding speed by 3.63$\times$ on CPU with translation quality comparable to that of an AR model.
\end{abstract}


\IEEEpeerreviewmaketitle

\section{Introduction}
\IEEEPARstart{B}{reaking} language barriers by using machines is an ultimate goal for international communications.
Automatic speech translation (ST) has been studied for this purpose for decades~\cite{stentiford1988machine,waibel1991janus,ney1999speech,fugen2009system}.
Cascade approaches combining automatic speech recognition (ASR) and machine translation (MT) systems have been the de facto standard, but the research paradigm is shifting to end-to-end speech translation (E2E-ST) models thanks to advances in deep learning~\cite{listen_and_translate,weiss2017sequence,berard2018end}.
End-to-end models have several attractive properties, such as avoiding ASR error propagation and low-latency decoding.
However, the translation quality of E2E models still lags behind that of cascade systems when additional resources are available, although the gap is closing~\cite{ansari2020findings,sperber-paulik-2020-speech,bentivogli-etal-2021-cascade}.

Various solutions have been proposed to bridge the quality gap between cascade and E2E models, such as multi-task learning with auxiliary tasks~\cite{weiss2017sequence,berard2018end}, pre-training~\cite{berard2018end,bansal-etal-2019-pre,wang2020bridging,wang-etal-2020-curriculum}, knowledge distillation~\cite{liu2019end,inaguma-etal-2021-source}, and semi-supervised training~\cite{jia2019leveraging,pino2019harnessing}.
Most studies on E2E systems have focused on {\ar} (AR) models, which generate a target sequence from left to right.
However, the decoding speed is not fast enough for real-world applications because of the incremental update of decoder states.
This slows the decoding speed, especially when the Transformer architecture~\cite{vaswani2017attention} is adopted because of the self-attention operation to all past tokens in the decoder at each generation step.

Recently, non-AR (NAR) models have attracted attention to increase the decoding speed by discarding the conditional dependency of outputs in AR models.
Gu~{\etal}~\cite{gu2018non} proposed the first single-step NAR MT model, but it sacrifices translation quality.
Various models have been proposed to address this issue; improved single-step NAR models~\cite{guo2019non,li-etal-2019-hint,wang2019non,wei-etal-2019-imitation,shao2020minimizing,liu2020task,ghazvininejad2020aligned,du2021order,qian-etal-2021-glancing}, iterative-refinement models~\cite{lee2018deterministic,ghazvininejad-etal-2019-mask,ghazvininejad2020semi,kasai2020non}, latent alignment models~\cite{sun2019fast,libovicky2018end,saharia-etal-2020-non,gu-kong-2021-fully}, insertion-based models~\cite{gu2019levenshtein,stern2019insertion,chan2019kermit}, and latent variable models~\cite{kaiser2018fast,ma-etal-2019-flowseq,shu2020latent,tu-etal-2020-engine}.
NAR models have been successfully extended to other tasks such as text-to-speech synthesis~\cite{oord2018parallel,ren2019fastspeech}, ASR~\cite{chan2020imputer,higuchi2020mask,higuchi2021improved} and ST~\cite{inaguma2021orthros,chuang-etal-2021-investigating}.

However, what differentiates the NAR ST task from other NAR tasks is \textit{many-to-many} non-monotonic mapping.
In other words, source inputs, even though they correspond to the same word sequence, can vary significantly depending on speaker attributes, speaking rate, recording conditions and so on.
These problems do not exist in text-based tasks because discrete tokens represent the input instead of continuous signals.
It is also challenging to determine the target length from the speech in advance because it includes many silent frames and the input length is much longer than that of text.

\begin{figure}[t]
    \centering
    \includegraphics[width=1.0\linewidth]{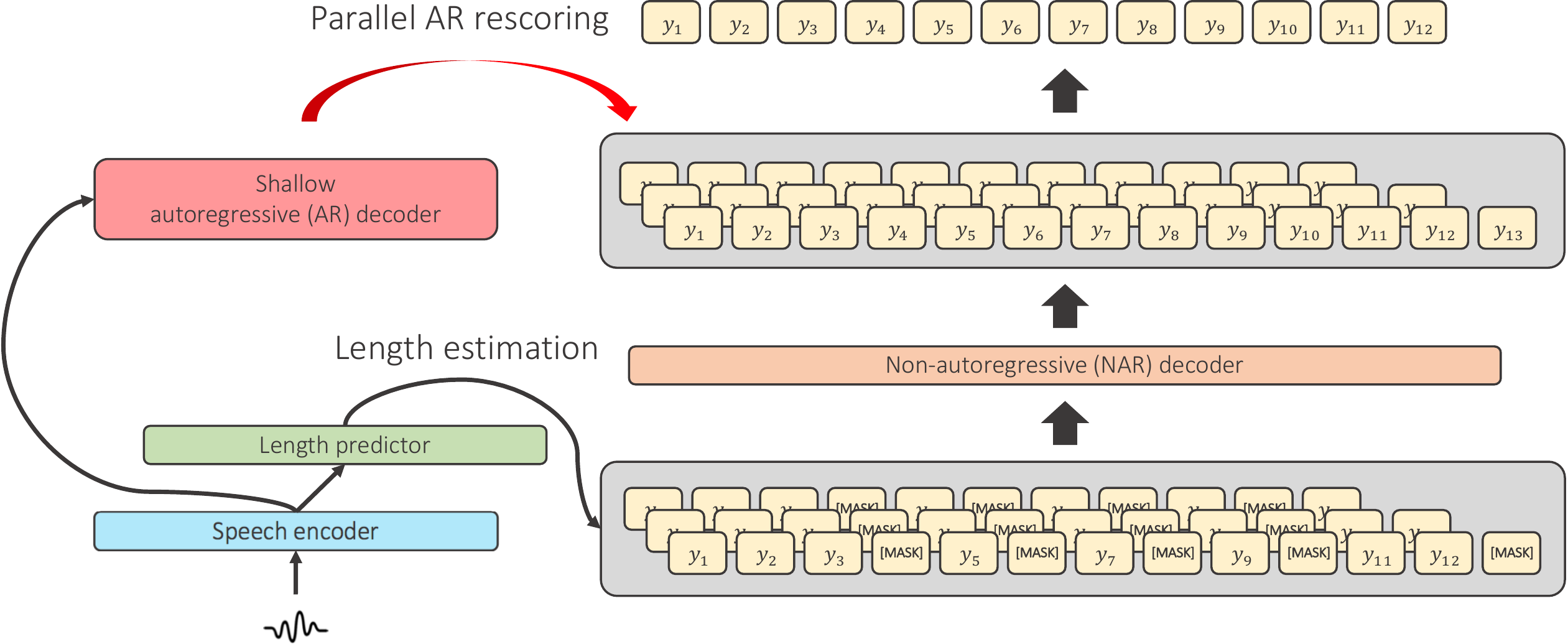}
    \caption{Overview of {\proposed}}
    \label{fig:orthros}
\end{figure}

To increase the decoding speed for the E2E-ST task, we propose an efficient unified NAR framework called \textit{{\proposed}}, which introduces an auxiliary shallow AR decoder on top of the shared speech encoder (see Fig.~\ref{fig:orthros}).
The AR decoder is jointly trained with the NAR decoder and is used to rescore multiple candidates of different lengths generated from the NAR decoder, referred to as \textit{parallel AR rescoring}.
This is based on an observation that sequence-level scores from the NAR decoder are not sufficient for accurately selecting the best candidate.
Because outputs from the NAR decoder can be fed to the AR decoder in parallel, the NAR decoding is still maintained.
Because a shallow AR decoder works for the rescoring purpose only, additional computation cost is minimal.

We investigate the conditional masked language model (CMLM)~\cite{ghazvininejad-etal-2019-mask} and connectionist temporal classification (CTC)-based model~\cite{ctc_graves} as NAR decoders for {\proposed}, referred to as {\proposed}-CMLM and {\proposed}-CTC, respectively.
The CMLM is an iterative refinement model while the CTC-based model is a single-step latent alignment model.
However, any other NAR decoder topology can be used for {\proposed} in theory as long as multiple candidates can be generated.
To enhance the training of the CMLM decoder, we further propose multi-mask training (MMT) and joint training with an auxiliary text-input NAR MT task.
With these methods, the CMLM decoder can take decoder inputs from multiple views per sample for improving translation quality.
To enhance the encoder representation, we also investigated the Conformer encoder~\cite{gulati2020}.

Experimental evaluations on three benchmark corpora, including six language directions, are conducted to show that {\proposed} significantly improves the BLEU scores with small additional latency.
We show that the rescoring is more effective than refining predictions through further iterations.
We also argue that joint training with the auxiliary shallow AR decoder improves the BLEU scores in most cases.
MMT and joint training with the NAR MT task also boosts the BLEU score for the Transformer encoder while maintaining the decoding speed.
The Conformer encoder improves BLEU scores across models, especially for CTC-based models.
We show that {\proposed}-CTC with the Conformer encoder achieved the best BLEU scores among NAR models.
Compared with the AR model, it achieved 3.63$\times$ faster decoding speed on a CPU with comparable BLEU scores.

The contributions of this article are summarized as follows:\footnote{This study is an extension of our previous study~\cite{inaguma2021orthros}, in which only the first item was investigated.}
\begin{itemize}
    \item We propose {\proposed} and show its effectiveness with two NAR models, {\ie}, CMLM and CTC-based model as examples.
    \item We propose two enhanced training methods for the CMLM.
    \item We thoroughly compare AR and NAR models on the basis of both Transformer and Conformer encoder architectures.
    \item We present strong AR baselines with and without sequence-level knowledge distillation (SeqKD)~\cite{kim-rush-2016-sequence}, which will be helpful for future E2E-ST studies.
    \item We conducted various analyses to evaluate the effectiveness of \textit{{\proposed}} in terms of model capacity, robustness against long-form speech, and searchability.
\end{itemize}

\section{Background}
In this section, we review AR and NAR E2E-ST models.
Let $\speech=(x_{1}, \ldots, x_{\xmax})$ denote input speech features in a source language, $\ytgt=(y_{1}, \ldots, y_{\ymax})$ denote the target translation text, where $\xmax$ and $\ymax$ denote the input length and output length, respectively.

\vspace{-2mm}
\subsection{AR sequence model}
Sequence-to-sequence models, so-called encoder-decoder models, are typically using AR sequence models such as the Transformer~\cite{vaswani2017attention}.
An AR model factorizes a conditional probability of $\ytgt$ given $\speech$ into a chain of conditional probabilities from left to right as follows:
\begin{eqnarray}
P(\ytgt|\speech)=\prod_{i=1}^{\ymax} \probar(y_{i} | y_{<i}, \speech), \label{eq:autoregressive}
\end{eqnarray}
where $\probar$ is a probability density function of the AR model, and $y_{<i}$ are all previous tokens before the $i$-th token.
The parameters are updated with a cross-entropy (CE) loss $\lossar$ formulated as
\begin{eqnarray*}
\lossar &=& - \log \probar(\ytgt|\speech) \\
&=& - \sum_{i=1}^{\ymax} \log \probar(y_{i} | y_{<i}, \speech).
\end{eqnarray*}
When using the Transformer architecture, training can be carried out efficiently by feeding all ground-truth tokens to the decoder in parallel, {\ie}, teacher-forcing~\cite{williams1989learning}.
During inference, however, beam search is used as a heuristic to find the most plausible sequence.
Because this incrementally expands a prefix of each hypothesis token by token, the decoding speed is typically not fast enough.
Specifically, the Transformer decoder slows the speed because it performs self-attention to all past tokens generated thus far.
The decoding complexity is ${\cal O}(\xmax \ymax)$.

\vspace{-2mm}
\subsection{NAR sequence model}
To increase the decoding of AR models by generating all tokens in a target sequence in parallel, the conditional independence for each token position in the output probability is assumed with NAR models.
The conditional probability in Eq.~\eqref{eq:autoregressive} is decomposed as
\begin{eqnarray}
P(\ytgt|\speech)=\prod_{i=1}^{\ymax} \probnar(y_{i} | \speech),  \label{eq:pure_nar}
\end{eqnarray}
where $\probnar$ is a probability density function of the NAR model.
Because $\probnar$ is not conditioned on $y_{<i}$, this formulation enables the generation of a sequence with a single iteration, which can achieve large increases in decoding speed.
However, such a strong assumption degrades translation quality because of the \textit{multimodality} problem, in which multiple correct translations are predicted given the same source sentence~\cite{gu2018non}.
SeqKD is an effective method for mitigating this problem by transforming reference translations in the training data into a more deterministic form by using a teacher AR model~\cite{zhou2019understanding}.

To relax the conditional independence assumption, iterative refinement methods~\cite{lee2018deterministic} have also been studied by modifying Eq.~\eqref{eq:pure_nar} to a chain of $\totaliter$ iterations as 
\begin{eqnarray*}
P(\ytgt | \speech) &=& \prod_{t=1}^{\totaliter} \probnar(Y^{\itert} | Y^{\itertminusone}, \speech) \\
&=& \prod_{t=1}^{\totaliter} \bigg( \prod_{i=1}^{\ymax^{\itert}} \probnar(y_{i}^{\itert} | Y^{\itertminusone}, \speech)\bigg),
\end{eqnarray*}
where $Y^{\itert}=(y_{1}^{\itert}, \cdots, y_{\ymax^{\itert}}^{\itert})$ is a sequence of tokens at the $t$-th iteration, where $\ymax^{\itert}$ is the output length at $t$-th iteration and can be changed in some models~\cite{gu2019levenshtein,stern2019insertion,chan2019kermit}.
Although iterative refinement methods slow the decoding speed of pure NAR models, they can achieve better translation quality in general and flexibly control the speed by changing $\totaliter$.

\vspace{2mm}
\subsubsection{CTC}
CTC is a loss function for a single-step latent alignment model, which eliminates the necessity of frame-level supervision for learning input-output mapping~\cite{ctc_graves}.
A linear projection layer is stacked on the encoder to generate a probability distribution $\probctc$ so the model is typically a decoder-free architecture.
The conditional independence per encoder frame is assumed with CTC, and CTC marginalizes posterior probabilities of all possible alignment paths efficiently with the forward-backward algorithm.
The gap in the sequence lengths between the input and output is filled by introducing blanks labels.
The CTC loss $\lossctc$ is defined as the negative log-likelihood:
\begin{eqnarray*}
\lossctc = - \log \probctc(\ytgt | \speech).
\end{eqnarray*}

During inference, CTC-based models can be regarded as fully NAR models if the best label class is taken from the vocabulary at every encoder frame, which is known as greedy search.
The decoding complexity is ${\cal O}(\xmax)$.
CTC assumes the length of $\speech$ to be longer than that of $\ytgt$ ({\ie}, $|\speech| \geq |\ytgt|$), so previous studies using CTC for the text-input NAR MT task adopted the upsampling technique to expand the input sequence length~\cite{libovicky2018end,saharia-etal-2020-non,gu-kong-2021-fully}.
However, this is not necessary for speech-to-text generation tasks because $|\speech|$ is generally much longer than $|\ytgt|$.
Instead, we usually compress input sequence lengths for the E2E-ST task~\cite{liu2020bridging,gaido-etal-2021-ctc,dong2021consecutive,zeng-etal-2021-realtrans}.
Another advantage of CTC for NAR models is that an explicit target length prediction is not necessary.
It can be obtained as a by-product after collapsing frame-level outputs.

\vspace{2mm}
\subsubsection{CMLM}
The CMLM~\cite{ghazvininejad-etal-2019-mask} is an iterative refinement-based model with a token in-filling adopted in Bidirectional Encoder Representations from Transformers (BERT)~\cite{devlin-etal-2019-bert}.
During inference, CMLM adopts the Mask-Predict algorithm, which alternates \textit{Mask} and \textit{Predict} steps at every iteration and refines predictions through $\totaliter$ iterations.
Therefore, the decoding complexity is ${\cal O}(\xmax \totaliter)$.

\vspace{2mm}
\noindent\textbf{Inference}\label{sssec:cmlm_inference}
Let $\hat{Y}_{\rm mask}^{\itert}$ and $\hat{Y}_{\rm obs}^{\itert}$ be masked and observed tokens in the prediction $\yhat^{\itert}$ at the $t$-th iteration ($0 \leq t \leq \totaliter$), respectively.
Given a target length $\ymaxhat$ estimated using a length predictor stacked on the encoder, the CMLM starts generation from a placeholder filled with a {\masktoken} token at all $\ymaxhat$ positions ($t=0$).
In the Predict step, the most plausible token is selected from the vocabulary according to the probability of the CMLM $\probcmlm$ at each masked position $i$ in $\hat{Y}_{\rm mask}^{\itertminusone}$ as
\begin{eqnarray*}
\hat{y}_{i}^{\itert} &=& \argmax_{w \in \vocab} \probcmlm(y_{i}=w | \hat{Y}_{\rm obs}^{\itert}, \speech), \\ 
p_{i, {\rm \textsc{cmlm}}}^{\itert} &\gets& \max_{w \in \vocab} \probcmlm(y_{i}=w | \hat{Y}_{\rm obs}^{\itert}, \speech),
\end{eqnarray*}
where $\vocab$ is the output vocabulary, and $p_{i, {\rm \textsc{cmlm}}}^{\itert}$ is a score of the $i$-th token at the $t$-th iteration.
Note that $p_{i, {\rm \textsc{cmlm}}}^{\itert}$ is updated for masked positions only.

In the Mask step, $k(t)$ tokens having the lowest confidence scores in the previous prediction $\hat{Y}^{\itert}$ are replaced with {\masktoken}, where $k(t)$ is a linear decay function defined as
\begin{eqnarray*}
k(t) &=& \lfloor \ymaxhat \cdot \frac{\totaliter - t}{t} \rfloor.
\end{eqnarray*}

As the target length must be determined before starting the token generation, length parallel decoding (LPD)~\cite{ghazvininejad-etal-2019-mask,wei-etal-2019-imitation} is typically used by generating multiple length candidates in parallel.
The $\lengthbeam$ length candidates are used by selecting top-$\lengthbeam$ classes from a length predictor.\footnote{We use top-$\lengthbeam$ classes instead of [$\hat{l} - \Delta$, $\hat{l} + \Delta$] centering on the best class $\hat{l}$ as in~\cite{wei-etal-2019-imitation}.}
After $\totaliter$ iterations, a candidate having the highest sequence-level scores $\frac{1}{\ymaxhat} \sum_{i} \log P_{i, {\rm \textsc{cmlm}}}^{\itert}$ is selected as the best translation output.

\vspace{2mm}
\noindent\textbf{Training}\label{sssec:cmlm_training}
The training objective of the CMLM is formulated as a CE loss calculated at masked positions as
\begin{eqnarray}
\losscmlm &=& - \sum_{y \in \ymask} \log \probcmlm(y|\yobs, \speech), \label{eq:objective_cmlm}
\end{eqnarray}
where $\ymask \subset \ytgt$ are partially masked ground-truth tokens, and $\yobs=\ytgt \setminus \ymask$.
The number of masked tokens is sampled from a uniform distribution $\mathcal{U}(1, \ymax)$, and the positions are also determined randomly.

A length predictor is implemented as a linear classifier and trained to predict the ground-truth target sequence length $\ymax$ given $\speech$ as
\begin{eqnarray*}
\losslp &=& - \log \problp(\ymax | \speech),
\end{eqnarray*}
where $\losslp$ is a length prediction loss.
Unlike the text-input MT task, where the encoder output corresponding to a spacial token {\lengthtoken} is used as an input to the linear classifier~\cite{ghazvininejad-etal-2019-mask}, time-averaged encoder outputs are used for the E2E-ST task.
The total objective $\losstotal$ is formulated as follows:
\begin{eqnarray*}
\losstotal &=& \losscmlm + \lambdalp \losslp,
\end{eqnarray*}
where $\lambdalp$ is a weight for the length-prediction loss.

\vspace{2mm}
\subsubsection{Semi-autoregressive Training}
Semi-AR training (SMART) is an improved training method to mitigate the gap in the behaviors of the CMLM between training and test~\cite{ghazvininejad2020semi}.
To resemble the test-time behavior based on the Mask-Predict algorithm during training, the decoder input is replaced with the model prediction at each training step.
To obtain the prediction, the most plausible token is taken at all $\ymax$ positions by adopting the same masking as the original CMLM training while the gradients are truncated.
Part of the resultant tokens $\yhat=(\hat{y}_{1}, \ldots, \hat{y}_{\ymax})$ are masked out to generate a new decoder input $\yhatobs$, which is fed to the decoder again to calculate the CE loss at \textit{all} positions as
\begin{eqnarray*}
\losscmlm=- \sum_{y \in \ytgt} \log \probcmlm(y | \yhatobs, \speech).
\end{eqnarray*}

During inference, the Mask-Predict algorithm is used, but tokens at all positions are updated at every iteration, unlike the original CMLM training.
Although SMART slightly slows the training speed by doubling the forward pass, it does not increase the decoding cost during inference.

\vspace{2mm}
\subsubsection{Mask-CTC}
Since the CMLM is designed for iterative refinement starting from {\masktoken} tokens at all positions, the translation quality at the early decoding iterations is likely to be poor.
To provide useful contexts from a single-step NAR E2E-ST model based on CTC, Mask-CTC initializes the decoder input of the CMLM with a greedy CTC prediction~\cite{higuchi2020mask}.
The CTC module is attached to the top encoder layer and trained jointly with the CMLM decoder.
The less confident CTC predictions, the probabilities of which are smaller than $p^{\rm thres}$, are replaced with {\masktoken} tokens.
However, unlike Mask-CTC for the ASR task, masked positions change depending on the score at each iteration because of the non-monotonic sequence generation in the ST task~\cite{higuchi2021improved}.
To keep from modifying most CTC outputs by using the original Mask-Predict algorithm, the restricted Mask-Predict algorithm was proposed~\cite{higuchi2021improved}, where $k(t)$ is truncated by the number of masked tokens at $t=0$.
We refer to this type of CMLM as \textit{CTC-CMLM} in this paper.

Because CTC can also be used as a length predictor, a separate length-prediction layer or LPD is not necessary.
The total training objective is formulated by interpolating $\losscmlm$ and $\lossctc$ as 
\begin{eqnarray*}
\losstotal=(1 - \lambdactc) \losscmlm + \lambdactc \lossctc,  \label{eq:total_loss_cmlmctc}
\end{eqnarray*}
where $\lambdactc$ is a CTC loss weight.

\subsection{Conformer}
The Conformer is a Transformer-based encoder architecture augmented by the convolution module in each block~\cite{gulati2020}.
It was introduced in the ASR task to capture global and local features by self-attention and convolution, respectively.
Because this property is compatible with speech, its effectiveness has been demonstrated in various speech-related tasks~\cite{guo2021recent}, including E2E-ST.
Each Conformer block introduces an additional convolution module right after the self-attention module.
An additional position-wise feed-forward network (FFN) is introduced right before the self-attention module, following the Macaron-Net~\cite{lu2019understanding}.
Relative positional encoding is also used in each self-attention module.
In our study, we investigated how the Conformer impacts the translation quality of both AR and NAR models.
Although the Conformer encoder introduces additional parameters, it is not a bottleneck of decoding speed.

\section{Proposed framework: {\proposed}}
In this section, we propose a unified NAR E2E-ST framework, \textit{{\proposed}}, which enhances the NAR decoder by joint training and parallel rescoring with an auxiliary shallow AR decoder.

\vspace{-2mm}
\subsection{Model architecture}
An overview of {\proposed} is presented in Fig.~\ref{fig:orthros}.
{\proposed} has three main components: speech encoder, NAR decoder, and auxiliary shallow AR decoder.
The motivation to introduce the AR decoder is based on an observation that sequence-level scores obtained from the NAR decoder are not suitable for selecting the best translation from multiple candidates.
This is because of the conditional independence assumption made with the NAR decoder, which makes the NAR decoder not fully leverage the effectiveness of generating multiple candidates.

In {\proposed}, the speech encoder is shared between the NAR and AR decoders, which greatly reduces the model size and computational overhead compared with using another AR model such as that proposed by Gu~{\etal}~\cite{gu2018non}.
For models except for a CTC-based one, a length predictor is also stacked on the speech encoder.
The entire architecture is jointly trained.

For {\proposed}, we mainly focus on the CMLM as the NAR decoder because we can control the decoding speed by changing the number of iterations, and it is widely used in the MT literature\cite{ghazvininejad-etal-2019-mask,ghazvininejad2020semi,guo-etal-2020-jointly,kong-etal-2020-incorporating,ding-etal-2020-context,xie-etal-2020-infusing,hao-etal-2021-multi,ghazvininejad2020aligned,du2021order,wang2021hybridregressive}.
We refer to this model as \textit{{\proposed}-CMLM}.
Because of powerful Conformer-CTC, we also investigated equipping the AR decoder of with CTC, referred to as \textit{{\proposed}-CTC}.
{\proposed} can be applied to any NAR decoder topology as long as multiple candidates can be generated.

\vspace{-2mm}
\subsection{Inference: parallel AR rescoring}
During inference, we use sequence-level scores from the AR decoder to select the most plausible translation after generating multiple candidates with the NAR decoder.
Because we can feed tokens at all positions to the AR decoder simultaneously in Eq.~\eqref{eq:autoregressive}, we refer to this as \textit{parallel AR rescoring}.

When using the CMLM as the NAR decoder, we use the Mask-Predict algorithm described in Section~\ref{sssec:cmlm_inference} to generate $\lengthbeam$ candidates through $\totaliter$ iterations.
When using a CTC-based model as the NAR decoder, we use left-to-right beam search with a beam width of $\lengthbeam$.
Although this is not purely NAR decoding, there is no matrix multiplication with weight parameters.
After the NAR decoding, we immediately feed the resulting tokens to the auxiliary AR decoder.
We use the sequence-level log probabilities $\frac{1}{\ymaxhat}\sum_{i} \log \probar(y_{i} | y_{<i}, \speech)$ obtained from the AR decoder to select the best translation among the $\lengthbeam$ candidates.
Note that scores from the NAR decoder are not used for the rescoring.\footnote{We also investigated linearly interpolating scores from the AR and NAR decoders in the log domain, but it did not lead to improvement.}

For {\proposed}-CMLM, this rescoring step corresponds to performing one more decoding iteration.
However, as shown in Section~\ref{sec:main_results}, parallel AR rescoring with $\totaliter-1$ is more effective than the Mask-Predict with $\totaliter$ without the rescoring.
A shallow AR decoder is used for rescoring purpose, so the additional decoding cost is much smaller than a single iteration of the CMLM decoder.

\vspace{-2mm}
\subsection{{\proposed}-CMLM}
In this section, we present \textit{{\proposed}-CMLM}, which uses the CMLM as the NAR decoder in {\proposed}. 
We also propose two enhanced training methods for the CMLM.

\vspace{2mm}
\subsubsection{MMT}
In the training of the CMLM, the decoder observes a single set of tokens per sample, and the CE loss is calculated at only masked positions.
This is not as data-efficient as AR models, especially when training data are sparse such as in the E2E-ST task.
To mitigate this problem, we introduce MMT by calculating CE losses for $\nmask$ forward passes with different $\yobs=\ytgt \setminus \ymask$.
At each forward pass, a random mask for $\ymask$ is sampled independently.
Accordingly, the CE loss in Eq.~\eqref{eq:objective_cmlm} is modified to the average CE losses in all forward passes as
\begin{eqnarray}
\losscmlm &=&  \frac{1}{\nmask} \sum_{m=1}^{\nmask} \losscmlm (\ymaskmmt | \yobsmmt, \speech)  \nonumber \\ 
&=& \frac{1}{\nmask} \sum_{m=1}^{\nmask} \sum_{y \in \ymaskmmt} - \log \probcmlm(y|\yobsmmt, \speech),  \label{eq:mmt}
\end{eqnarray}
where $\ymaskmmt$ and $\yobsmmt$ are masked and observed tokens in the $m$-th forward pass, respectively. 
This method corresponds to augmenting the training data by a factor of $\nmask$ in total.
However, the number of parameter updates does not increase.

\vspace{2mm}
\subsubsection{Auxiliary NAR MT objective}
To further enhance the CMLM decoder further, we train the NAR E2E-ST model jointly with an auxiliary text-input NAR MT task by sharing the parameters of the CMLM decoder.
The effectiveness of an auxiliary MT task has been studied to improve the translation quality of AR E2E-ST models by leveraging source transcriptions $\ysrc$~\cite{berard2018end,bahar2019comparative,liu2020bridging}.
However, the task is still AR sequence generation, and none of the previous studies investigated the effectiveness of the auxiliary MT task for the NAR E2E-ST task.
The CE loss for the NAR MT task $\lossmt$ is formulated as
\begin{eqnarray}
\lossmt=\frac{1}{\nmask} \sum_{m=1}^{\nmask} \sum_{y \in \ymaskmmt} - \log \probcmlm(y | \yobsmmt, \ysrc).
\end{eqnarray}
To encode source transcriptions, an additional text encoder is introduced.
However, this encoder can be removed during inference, so the decoding cost does not change.

Regarding masking strategies, we use separate masks for the NAR MT task by randomly sampling them independently, which is more effective than reusing the same mask used in the NAR E2E-ST task.
This also has an effect of augmenting the training data twice similar to MMT.
However, we show that these two methods are complementary and their combination\footnote{This corresponds to augmenting the training data by a factor of 4.} is effective.

\vspace{2mm}
\subsubsection{Training objective}
We optimize an entire network with an end-to-end training objective as
\begin{eqnarray}
\losstotal=\losscmlm + \lambdalp \losslp + \lambdaar \lossar + \lambdamt \lossmt,  \label{eq:total_loss_orthros_cmlm}
\end{eqnarray}
where $\lambda_{*}$ are the corresponding loss weights.\footnote{Unlike our previous study~\cite{inaguma2021orthros}, we removed the auxiliary CTC-based ASR objective because it can be removed without quality degradation when pre-training the encoder with the ASR task. Instead, an auxiliary NAR MT task was newly introduced in this article.}

\vspace{-2mm}
\subsection{{\proposed}-CTC}
In this section, we describe \textit{{\proposed}-CTC}, which uses a CTC-based model as the NAR decoder.
Unlike {\proposed}-CMLM, {\proposed}-CTC does not have any NAR decoder parameters except for the linear projection layer, so it is more parameter-efficient.
Note again that CTC does not require a separate length predictor.
As discussed in Section~\ref{sec:main_results}, translation quality of CTC-based models greatly improves using the Conformer encoder.
Therefore, we focus only on the Conformer encoder.
The total objective $\losstotal$ is formulated as
\begin{eqnarray}
\losstotal=\lossctc + \lambdaar \lossar. \label{eq:total_loss_orthros_ctc}
\end{eqnarray}

To generate multiple candidates, left-to-right prefix beam search can be used in the CTC branch.
Although this is no longer NAR decoding, we can use an efficient implementation with C++\footnote{\url{https://github.com/parlance/ctcdecode}} as done in a previous study~\cite{gu-kong-2021-fully} because there is no matrix multiplication.

\section{Experimental setting}\label{sec:experiment}

\subsection{Dataset}
We used Must-C~\cite{di-gangi-etal-2019-must} En$\to$\{De, Nl, Fr, Es, It\}, Libri-trans En$\to$Fr~\cite{libri_trans}, and Fisher-CallHome Spanish Es$\to$En~\cite{fisher_callhome} corpora for the experimental evaluations.
We used the recipes provided in the ESPnet-ST toolkit~\cite{inaguma-etal-2020-espnet}.

\vspace{2mm}
\subsubsection{Must-C En$\to$\{De, Nl, Fr, Es, It\}}
We used En$\to$De (399 h), En$\to$Nl (433 h), En$\to$Fr (483 h), En$\to$Es (495 h), and En$\to$It (456 h) directions on Must-C.
This corpus contains spontaneous English lecture speech extracted from TED talks, the corresponding source transcriptions, and the target translations.
Non-verbal speech labels, such as ``\textit{(Applause)}'' and  ``\textit{(Laughter)}'', were kept during training but removed during evaluation as post-processing.
We report case-sensitive detokenized BLEU scores~\cite{papineni-etal-2002-bleu} on the \texttt{tst-COMMON} set.

\vspace{2mm}
\subsubsection{Libri-trans (En$\to$Fr)}
This corpus contains 100 h of English read speech, the corresponding English transcriptions, and the French translations.
The French translation in the training set is augmented with Google Translate for each utterance, and we used both references following the standard practice.
We report case-insensitive detokenized BLEU scores on the \texttt{test} set.

\vspace{2mm}
\subsubsection{Fisher-CallHome Spanish (Es$\to$En)}
This corpus contains 171 h of Spanish conversational telephone speech, the corresponding Spanish transcriptions as well as the English translations.
Following the standard practice of this corpus, all punctuation marks except for apostrophes were removed from both transcriptions and translations.
We report case-insensitive BLEU scores on Fisher-\{\texttt{dev}, \texttt{dev2}, \texttt{test}\}, and CallHome-\{\texttt{devtest}, \texttt{evltest}\}.
We report case-insensitive detokenized BLEU scores on these five sets.
Note that BLEU scores on the Fisher splits were evaluated with four references.
We used the Fisher-\texttt{dev} set as the validation set.

\vspace{-1mm}
\subsection{Pre-processing}
We extracted 80-channel log-mel filterbank coefficients computed with 25-ms window size and shifted every 10ms with three-dimensional pitch features with the Kaldi toolkit~\cite{kaldi}.
We augmented speech data with speed perturbation~\cite{speed_perturbation} and SpecAugment~\cite{specaugment} for both ASR and E2E-ST task.
Utterances having more than 3000 speech frames or more than 400 characters were removed from the training data to fit the GPU memory.

We tokenized all sentences with the {\tt tokenizer.perl} script in the Moses toolkit~\cite{koehn-etal-2007-moses}.
Source transcriptions were lowercased, and the punctuation marks except for apostrophes were removed.
We constructed shared source and target vocabularies on the basis of the byte pair encoding (BPE) algorithm~\cite{sennrich-etal-2016-neural} with the Sentencepiece toolkit~\cite{kudo-richardson-2018-sentencepiece}.
ASR vocabularies were built only on source transcriptions.
For AR models, we used 1k units for all tasks on the Fisher-CallHome Spanish and Libri-trans corpora, while 5k and 8k units were used for ASR and E2E-ST/MT tasks on the Must-C corpus, respectively.
For NAR models, we used 16k units on all corpora, unless otherwise noted.
These vocabulary sizes were selected to achieve the best performance for each model following~\cite{inaguma2021orthros}.

\begin{table*}[t]
    \centering
    \begingroup
    \caption{BLEU scores with \textbf{Transformer encoder} on \underline{Must-C} \texttt{tst-COMMON}. Decoding speed was measured on En$\to$De with batch size of 1. $\ndecar$ was set to 1. MMT: multi-mask training. NAR MT: auxiliary text-input NAR MT objective.}
    \label{tab:taslp2021_results_mustc_transformer}
    \vspace{-1mm}
    \scalebox{1.0}{
    \begin{tabular}{@{}clccccccccc@{}}\toprule
      \multicolumn{2}{c}{\multirow{2}{*}{Model}} & \multirow{2}{*}{$\totaliter$} & \multicolumn{6}{c}{BLEU ($\uparrow$)} & \multicolumn{2}{c}{Speedup ($\uparrow$)} \\
      \cmidrule(lr){4-9} \cmidrule(lr){10-11}
      \multicolumn{2}{c}{} &  & De & Nl & Fr & Es & It & Avg & GPU & CPU \\
      \midrule
      
            \multirow{7}{*}{\shortstack{E2E AR}} & \ ESPnet-ST~\cite{inaguma-etal-2020-espnet} ($\beamst=10$) & \multirow{3}{*}{$\ymax$} & 22.9 & 27.4 & 32.8 & 28.0 & 23.8 & 27.0 & -- & -- \\
            & \ Fairseq S2T~\cite{wang-etal-2020-fairseq} &  & 22.7 &  27.3 & 32.9 & 27.2 & 22.7 & 26.6 & -- & -- \\
            & \ NeurST~\cite{zhao-etal-2021-neurst} & & 22.8 &  27.2 & 33.3 & 27.4 & 22.9 & 26.7 & -- & -- \\
            \cdashlinelr{2-11}
            
            & \ Transformer ($\beamst=1$) & \multirow{4}{*}{$\ymax$} & 21.0 & 26.1 & 31.6 & 26.2 & 22.1 & 25.4 & \phantom{0}1.64$\times$ & \phantom{0}3.26$\times$ \\
            & \ Transformer ($\beamst=4$) &  & 22.8 & 27.3 & 33.3 & 27.8 & 23.3 & 26.9 & \phantom{0}1.00$\times$ & \phantom{0}1.00$\times$ \\
            & \ Transformer + SeqKD ($\beamst=1$) &  & 23.8 & 27.7 & 33.7 & 28.0 & 23.3 & 27.3 & \phantom{0}1.64$\times$ & \phantom{0}3.26$\times$ \\
            & \ Transformer + SeqKD ($\beamst=4$) &  & \bf{24.3} & \bf{28.4} & \bf{34.5} & \bf{28.9} & \bf{24.2} & \bf{28.1} & \phantom{0}1.00$\times$ & \phantom{0}1.00$\times$ \\
            \cmidrule{1-11}
            
            \multirow{10}{*}{\shortstack{E2E NAR}} 
            & \ CTC & 1 & 19.8 & 23.9 & 28.9 & 22.8 & 19.5 & 23.0 & 24.51$\times$ & 14.62$\times$ \\
             \cdashlinelr{2-11}
            
            & \ CMLM & \multirow{4}{*}{4} & 20.1 & 23.6 & 28.9 & 24.0 & 20.5 & 23.4 & \phantom{0}6.45$\times$ & \phantom{0}5.43$\times$ \\
            & \ SMART &  & 20.4 & 24.3 & 29.7 & 24.9 & 21.0 & 24.1 & \phantom{0}6.45$\times$ & \phantom{0}5.43$\times$ \\
            & \ {\proposed}-CMLM &  & 21.8 & 25.3 & 30.0 & 25.3 & 21.5 & 24.8 & \phantom{0}5.69$\times$ & \phantom{0}5.10$\times$ \\
            & \ \ + MMT + NAR MT & & \bf{22.5} & \bf{26.1} & \bf{31.5} & \bf{26.0} & \bf{22.7} & \bf{25.8} & \phantom{0}5.69$\times$ & \phantom{0}5.10$\times$ \\
            \cdashlinelr{2-11}
            
            & \ CMLM & \multirow{4}{*}{10} & 21.5 & 25.1 & 30.7 & 25.3 & 21.8 & 24.9 & \phantom{0}3.13$\times$ & \phantom{0}2.75$\times$ \\
            & \ SMART &  & 21.4 & 25.2 & 31.0 & 25.4 & 22.1 & 25.0 & \phantom{0}3.13$\times$ & \phantom{0}2.75$\times$ \\
            & \ {\proposed}-CMLM &  & 22.9 & 26.4 & 31.8 & 26.1 & 22.5 & 26.0 & \phantom{0}2.99$\times$ & \phantom{0}2.64$\times$ \\
            & \ \ + MMT + NAR MT & & \bf{23.5} & \bf{27.1} & \bf{33.1} & \bf{27.1} & \bf{23.3} & \bf{26.8} & \phantom{0}2.99$\times$ & \phantom{0}2.64$\times$ \\
            \cmidrule{1-11}
            
        Cascade AR 
            & \ Trf ASR $\to$ Trf MT & $2\ymax$ & 23.5 & 28.5 & 33.9 & 28.6 & 24.3 & 27.8 & \phantom{0}0.45$\times$ & \phantom{0}0.68$\times$ \\
            \bottomrule
    \end{tabular}
    }
    \endgroup
\end{table*}

\begin{table*}[t]
    \centering
    \begingroup
    \caption{BLEU scores with \textbf{Conformer encoder} on \underline{Must-C} \texttt{tst-COMMON}}
    \label{tab:taslp2021_results_mustc_conformer}
    \vspace{-1mm}
    \scalebox{1.0}{
    \begin{tabular}{@{}clccccccccc@{}}\toprule
      \multicolumn{2}{c}{\multirow{2}{*}{Model}} & \multirow{2}{*}{$\totaliter$} & \multicolumn{6}{c}{BLEU ($\uparrow$)} & \multicolumn{2}{c}{Speedup ($\uparrow$)} \\
      \cmidrule(lr){4-9} \cmidrule(lr){10-11}
      \multicolumn{2}{c}{} &  & De & Nl & Fr & Es & It & Avg & GPU & CPU \\
      \midrule
            \multirow{4}{*}{\shortstack{E2E AR}} & \ Conformer ($\beamst=1$) & \multirow{4}{*}{$\ymax$} & 23.2 & 28.3 & 34.5 & 29.7 & 24.6 & 28.1 & \phantom{0}1.59$\times$ & \phantom{0}2.82$\times$ \\
            & \ Conformer ($\beamst=4$) &  & 25.0 & 29.7 & 35.5 & 30.5 & 25.4 & 29.2 & \phantom{0}1.00$\times$ & \phantom{0}1.00$\times$ \\
            & \ Conformer + SeqKD ($\beamst=1$) &  & 25.7 & 29.7 & 35.8 & 30.6 & 25.5 & 29.5 & \phantom{0}1.59$\times$ & \phantom{0}2.82$\times$ \\
            & \ Conformer + SeqKD ($\beamst=4$) &  & \bf{26.3} & \bf{30.6} & \bf{36.4} & \bf{31.0} & \bf{25.9} & \bf{30.0} & \phantom{0}1.00$\times$ & \phantom{0}1.00$\times$ \\
            \cmidrule{1-11}

            \multirow{12}{*}{\shortstack{E2E NAR}} & \ CTC ($\lengthbeam=1$) & \multirow{2}{*}{1} & 24.1 & 28.5 & 34.6 & 29.0 & 24.3 & 28.1 & 13.83$\times$ & \phantom{0}8.32$\times$ \\
            & \ {\proposed}-CTC ($\lengthbeam=20$) &  & \bf{25.3} & \bf{29.9} & \bf{36.2} & \bf{30.4} & \bf{25.4} & \bf{29.4} & \phantom{0}1.14$\times$ & \phantom{0}3.63$\times$ \\  %
            \cdashlinelr{2-11}
            
            & \ CMLM & \multirow{4}{*}{4} & 22.6 & 26.3 & 31.3 & 26.5 & 21.6 & 25.7 & \phantom{0}5.44$\times$ & \phantom{0}4.20$\times$ \\
            & \ CTC-CMLM &  & 24.1 & 27.8 & 34.7 & 29.1 & 24.5 & 28.0 & \phantom{0}5.07$\times$ & \phantom{0}5.97$\times$ \\
            & \ {\proposed}-CMLM &  & \bf{23.4} & \bf{27.6} & 33.3 & 28.3 & 23.2 & 27.2 & \phantom{0}4.92$\times$ & \phantom{0}4.03$\times$ \\
            & \ \ + MMT + NAR MT & & 23.3 & 27.5 & \bf{34.0} & \bf{28.7} & \bf{23.7} & \bf{27.4} & \phantom{0}4.92$\times$ & \phantom{0}4.03$\times$\\
            \cdashlinelr{2-11}

            & \ CMLM & \multirow{4}{*}{10} & 23.5 & 27.6 & 33.0 & 27.5 & 22.7 & 26.9 & \phantom{0}2.89$\times$ & \phantom{0}2.33$\times$ \\
            & \ CTC-CMLM &  & 24.2 & 28.2 & 34.8 & 29.3 & 24.6 & 28.1 & \phantom{0}3.06$\times$ & \phantom{0}4.43$\times$ \\

            & \ {\proposed}-CMLM &  & \bf{24.5} & 28.5 & 34.6 & 29.1 & 24.0 & 28.1 & \phantom{0}2.73$\times$ & \phantom{0}2.31$\times$ \\
            & \ \ + MMT + NAR MT &  & 24.1 & \bf{28.6} & \bf{35.1} & \bf{29.2} & \bf{24.4} & \bf{28.3} & \phantom{0}2.73$\times$ & \phantom{0}2.31$\times$ \\
            \midrule
            
        Cascade AR 
            & \ Cfm ASR $\to$ Trf MT & $2\ymax$ & 24.1 & 29.4 & 35.0 & 29.5 & 24.7 & 28.5 & \phantom{0}0.46$\times$ & \phantom{0}0.67$\times$ \\
            \bottomrule
    \end{tabular}
    }
    \endgroup
\end{table*}

\subsection{Architecture}
We implemented models based on the ESPnet-ST toolkit.
All models in the ASR and E2E-ST tasks consisted of 12 encoder blocks and 6 decoder blocks, while MT models used 6 Transformer encoder blocks.
We also used the Transformer decoder when using the Conformer encoder.
The speech encoder in the ASR and E2E-ST tasks had two convolutional neural network (CNN) blocks before the first encoder block, which had a kernel size of 3 and channel size of 256.
Each CNN block down-sampled features on both time and frequent axes with a stride of two, which resulted in the four-fold time reduction.
The dimensions of the self-attention layer $\dmodel$ and FFN $\dff$ were set to 256 and 2048, respectively, and the number of attention heads $\nhead$ was set to four.
The kernel size of depthwise separable convolution in each Conformer block was set to 15.
The CTC-based models had the same encoder architecture while the decoder was replaced with a linear projection layer.

\subsection{Training}
The Adam optimizer~\cite{adam} was used for training with $\beta_{1}=0.9$, $\beta_{2}=0.98$, and $\epsilon=10^{-9}$.
We used the Noam learning rate schedule~\cite{vaswani2017attention} with warmup steps of 25k and learning-rate constant of 5.0.
The effective batch size was set to 256 utterances for the NAR models.
We used dropout and label smoothing~\cite{label_smoothing} with a probability of 0.1 and 0.1, respectively.
We set ($\lambdalp$, $\lambdaar$, $\lambdamt$) to (0.1, 0.3, 0.3) in Eq.~\eqref{eq:total_loss_orthros_cmlm} throughout the experiments and empirically confirmed that they work well under various conditions.
The number of masks for MMT $\nmask$ was set to 2.
We trained AR models for 30 epochs, CTC-based models and CTC-CMLM for 100 epochs, the other models for 50 epochs.
The last five best checkpoints based on the validation score were used for model averaging, except that the last ten best checkpoints were used for the CTC-based models.

We initialized encoder parameters of all E2E-ST models with those of the pre-trained ASR model trained on the same speech data.
We did not use any external resources for pre-training.
The decoder parameters of all E2E-ST models were initialized on the basis of a strategy in BERT~\cite{devlin-etal-2019-bert}, where weight parameters were sampled from $N(0, 0.02)$, biases were set to zero, and layer normalization parameters were set to $\beta=0$, $\gamma=1$.\footnote{Unlike~\cite{inaguma2021orthros}, we did not use a pre-trained MT model for initialization of decoder parameters.}
This technique was used for the CMLM models in the MT task~\cite{ghazvininejad-etal-2019-mask}.

Following the standard practice in NAR models~\cite{gu2018non,ghazvininejad-etal-2019-mask}, we used SeqKD with an AR Transformer MT model as a teacher, except for Libri-trans.\footnote{We did not observe any improvement in BLEU scores with SeqKD for both AR and NAR models on this corpus. This is probably because the teacher MT model was too weak due to a small amount of the training bitext.}
The teacher MT models used a beam width of 5.

\subsection{Decoding}
For the AR models, we used a beam width $\beamst \in \{1, 4, 10\}$.
The ASR models used the joint CTC/Attention decoding~\cite{hybrid_ctc_attention} with shallow fusion of an external long short-term memory (LSTM) LM.
For the CMLM decoders, we used a length beam size $\lengthbeam=5$ and number of iterations $\totaliter \in \{4, 10\}$ as default settings.
We mainly used a relatively small $\lengthbeam$ to avoid slowing the decoding speed on the CPU.
We also used a de-duplication technique~\cite{lee2018deterministic,sun2020approach,du2021order} for the CMLM decoders, where repeated tokens were collapsed to a single token as postprocessing, except for the Fisher-CallHome Spanish corpus.\footnote{De-duplication was effective for corpora having long-form speech, such as Must-C and Libri-trans.}
For {\proposed}-CTC, we used $\lengthbeam=20$ without de-duplication.
The decoding speed was measured with a batch size of 1 on an NVIDIA TITAN RTX GPU and Intel(R) Xeon(R) Gold 6128 CPU @ 3.4GHz by averaging on five runs.
We calculated BLEU scores with SacreBLEU\footnote{case.mixed+numrefs.1+smooth.exp+tok.13a+version.1.5.1}~\cite{post-2018-call}.

\section{Main results}\label{sec:main_results}
\subsection{Must-C}

\vspace{2mm}
\subsubsection{Transformer encoder}
Results with the Transformer encoder on Must-C are shown in Table~\ref{tab:taslp2021_results_mustc_transformer}.
The single step NAR decoding with CTC showed poor BLEU scores although the speed increases were very large; 24.51$\times$ and 14.62$\times$ on the GPU and CPU, respectively.
The iterative NAR models, however, improved BLEU scores at the cost of speed.
We first compared the CMLM and SMART without the auxiliary AR decoder, and the SMART outperformed the CMLM when $\totaliter=4$.
However, the gains were diminished when $\totaliter=10$.
Therefore, we did not use SMART for {\proposed}.
{\proposed}-CMLM consistently improved the CMLM for all language pairs regardless of $\totaliter$.
The additional latency with the parallel AR rescoring was about 5 and 100 ms on the GPU and CPU, respectively, which are negligible when we take the gains of BLEU scores into account.
The enhanced training with MMT and multi-task with the auxiliary NAR-MT task further improved the BLEU scores for all language pairs and $\totaliter$.
Compared with the baseline CMLM, the enhanced {\proposed}-CMLM brought the gains of 2.4 and 1.9 BLEU scores on average for $\totaliter=4$ and $\totaliter=10$, respectively.
The enhanced {\proposed}-CMLM with $\totaliter=10$ also achieved BLEU scores comparable to those of the AR Transformer ($\beamst=4$) with 2.99$\times$ and 2.64$\times$ increase in speed on the GPU and CPU, respectively.
However, we found that SeqKD improved the AR model and the gains were much larger than those reported in the MT literature~\cite{gu-kong-2021-fully}.
Compared with the cascade system, the enhanced {\proposed}-CMLM with $\totaliter=10$ achieved 6.44$\times$ and 3.88$\times$ faster decoding on GPU and CPU while the average BLEU score underperformed by 1.0.

\vspace{2mm}
\subsubsection{Conformer encoder}
Results with the Conformer encoder on Must-C are shown in Table~\ref{tab:taslp2021_results_mustc_conformer}.
We observed that the Conformer encoder consistently improved BLEU scores across models.
The CTC-based models showed the largest gains of BLEU scores by more than 4 BLEU.
One limitation of the CTC-based models was that it took more training steps to converge.
We trained them for 100 epochs while we trained the CMLM models for 50 epochs.
We evaluated {\proposed}-CTC considering the strong BLEU scores of Conformer-CTC.
It resulted in further gains by parallel AR rescoring and addressed the slow convergence issue.
In terms of decoding speed, {\proposed}-CTC achieved a large gain for the CPU (3.63$\times$) while that for the GPU was small (1.14$\times$).
We also investigated CTC-CMLM, which refined a CTC output through multiple iterations.
Although it outperformed {\proposed}-CMLM in speed with similar BLEU scores, we did not observe any BLEU improvement from the pure CTC-based model.
Therefore, increasing the number of candidates was more effective than refining the single output multiple times.
Similar to the Transformer encoder, {\proposed}-CMLM improved the CMLM by 1.5 and 1.2 BLEU scores on average for $\totaliter=4$ and $\totaliter=10$, respectively.
The enhanced training slightly improved the BLEU scores (+0.2) except for En$\to$De and En$\to$Nl.
We reason that the Conformer encoder already enhanced the CMLM decoder by providing better encoder representations and augmenting masks for the CMLM training was not complementary.
Comparing the enhanced {\proposed}-CMLM ($\totaliter=10$) with the cascade system, the gap in the average BLEU score was reduced from 1.0 to 0.2 by using the Conformer encoder.
To summarize, {\proposed}-CTC showed the best BLEU scores in all language pairs.

\begin{table}[t]
    \centering
    \tabcolsep 2pt
    \begingroup
    \caption{BLEU scores with \textbf{Transformer encoder} on \underline{Libri-trans En$\to$Fr} \texttt{test}. $\ndecar$ was set to 1.}
    \label{tab:taslp2021_results_libritrans_transformer}
    \vspace{-1mm}
    \scalebox{1.0}{
    \begin{tabular}{@{}clcccc@{}}\toprule
      \multicolumn{2}{c}{\multirow{2}{*}{Model}} & \multirow{2}{*}{$\totaliter$} & \multirow{2}{*}{BLEU ($\uparrow$)} & \multicolumn{2}{c}{Speedup ($\uparrow$)} \\ \cmidrule(lr){5-6}
      \multicolumn{2}{c}{} &  & & GPU & CPU \\ \midrule
            \multirow{10}{*}{\shortstack{E2E AR}} & \ WordKD~\cite{liu2019end} & \multirow{8}{*}{$\ymax$} & 17.02 & -- & -- \\
            & \ TCEN-LSTM~\cite{wang2020bridging} &  & 17.05 & -- & -- \\
            & \ NeurST~\cite{zhao-etal-2021-neurst} & & 17.2\phantom{0} & -- & -- \\
            & \ Curriculum PT~\cite{wang-etal-2020-curriculum} &  & 17.66 & -- & -- \\
            & \ LUT~\cite{dong2021listen} &  & 17.75 & -- & -- \\
            & \ STAST~\cite{liu2020bridging} &  & 17.81 & -- & -- \\
            & \ COSTT~\cite{dong2021consecutive} &  & 17.83 & -- & -- \\
            & \ SATE~\cite{xu-etal-2021-stacked} &  & 18.3\phantom{0} & -- & -- \\
            \cdashlinelr{2-6}
            
            & \ Transformer ($\beamst=1$) & \multirow{2}{*}{$\ymax$} & 16.6\phantom{0} & \phantom{0}1.64$\times$ & \phantom{0}4.29$\times$ \\ 
            & \ Transformer ($\beamst=4$) &  & 16.7\phantom{0} & \phantom{0}1.00$\times$ & \phantom{0}1.00$\times$ \\
            \midrule
            
            \multirow{10}{*}{\shortstack{E2E NAR}}
            & \ CTC ($\lengthbeam=1$) & 1 & 13.4\phantom{0} & 35.03$\times$ & 26.61$\times$ \\  %
            \cdashlinelr{2-6}
            
            & \ CMLM & \multirow{4}{*}{4} & 14.3\phantom{0} & \phantom{0}9.40$\times$ & 10.09$\times$ \\
            & \ SMART &  & 13.6\phantom{0} & \phantom{0}9.40$\times$ & 10.09$\times$ \\
            & \ {\proposed}-CMLM &  & 15.0\phantom{0} & \phantom{0}8.47$\times$ & \phantom{0}9.52$\times$ \\
            & \ \ + MMT + NAR MT &  & \bf{15.9}\phantom{0} & \phantom{0}8.47$\times$ & \phantom{0}9.52$\times$ \\
            \cdashlinelr{2-6}

            & \ CMLM & \multirow{4}{*}{10} & 15.5\phantom{0} & \phantom{0}4.61$\times$ & \phantom{0}5.18$\times$ \\
            & \ SMART &  & 14.6\phantom{0} & \phantom{0}4.61$\times$ & \phantom{0}5.18$\times$ \\
            & \ {\proposed}-CMLM &  & 16.2\phantom{0} & \phantom{0}4.39$\times$ & \phantom{0}4.99$\times$ \\
            & \ \ + MMT + NAR MT &  & \bf{16.7}\phantom{0} & \phantom{0}4.39$\times$ & \phantom{0}4.99$\times$ \\
            \midrule
            
        Cascade AR
            & Trf ASR $\to$ Trf MT & $2\ymax$ & 17.0\phantom{0} & \phantom{0}0.43$\times$ & \phantom{0}0.77$\times$ \\
        \bottomrule
    \end{tabular}
    }
    \endgroup
\end{table}

\subsection{Libri-trans}
The results from the Transformer and Conformer encoders on Libri-trans are listed in Tables~\ref{tab:taslp2021_results_libritrans_transformer} and \ref{tab:taslp2021_results_libritrans_conformer}, respectively.
We did not use SeqKD for this corpus because it was not effective in our preliminary experiments.
We also referred to the results of previous studies that did not leverage additional resources.

For Transformer-based models, we confirmed that SMART was not effective and {\proposed} outperformed both CMLM and SMART regardless of $\totaliter$.
The enhanced training was also helpful, boosting the BLEU score by 0.9 for $\totaliter=4$ and 0.5 for $\totaliter=10$.
The enhanced {\proposed} with $\totaliter=10$ matched the AR model in terms of BLEU score with 4.39$\times$ and 4.99$\times$ increase in decoding speed on the GPU and CPU, respectively.
The relative increase in speed were much larger than those on Must-C because the \texttt{test} set on Libri-trans had many long-form utterances up to about 50 s.

\begin{table}[t]
    \centering
    \tabcolsep 2pt
    \begingroup
    \caption{BLEU scores with \textbf{Conformer encoder} on \underline{Libri-trans En$\to$Fr} \texttt{test}}
    \label{tab:taslp2021_results_libritrans_conformer}
    \vspace{-1mm}
    \scalebox{1.0}{
    \begin{tabular}{@{}clcccc@{}}\toprule
      \multicolumn{2}{c}{\multirow{2}{*}{Model}} & \multirow{2}{*}{$\totaliter$} & \multirow{2}{*}{BLEU ($\uparrow$)} & \multicolumn{2}{c}{Speedup ($\uparrow$)} \\ \cmidrule(lr){5-6}
      \multicolumn{2}{c}{} &  & & GPU & CPU \\ \midrule
            \multirow{2}{*}{\shortstack{E2E AR}}
            & \ Conformer ($\beamst=1$) & \multirow{2}{*}{$\ymax$} & 18.8\phantom{0} & \phantom{0}1.61$\times$ & \phantom{0}3.90$\times$ \\
            & \ Conformer ($\beamst=4$) &  & \bf{19.2}\phantom{0} & \phantom{0}1.00$\times$ & \phantom{0}1.00$\times$ \\
            \midrule

            \multirow{10}{*}{\shortstack{E2E NAR}} 
            & \ CTC ($\lengthbeam=1$) & \multirow{2}{*}{1} & 17.5\phantom{0} & 19.24$\times$ & 15.53$\times$ \\
            & \ {\proposed}-CTC ($\lengthbeam=20$) &  & \bf{18.5}\phantom{0} & \phantom{0}1.37$\times$ & \phantom{0}3.52$\times$ \\  %
           \cdashlinelr{2-6}
        
            & \ CMLM & \multirow{4}{*}{4} & 15.3\phantom{0} & \phantom{0}7.92$\times$ & \phantom{0}8.23$\times$ \\
            & \ CTC-CMLM &  & \bf{17.0}\phantom{0} & \phantom{0}7.03$\times$ & 11.34$\times$ \\
            & \ {\proposed}-CMLM &  & 16.7\phantom{0} & \phantom{0}7.11$\times$ & \phantom{0}7.82$\times$ \\
            & \ \ + MMT + NAR MT &  & 16.7\phantom{0} & \phantom{0}7.11$\times$ & \phantom{0}7.82$\times$ \\
            \cdashlinelr{2-6}

            & \ CMLM & \multirow{4}{*}{10} & 15.8\phantom{0} & \phantom{0}4.32$\times$ & \phantom{0}4.77$\times$ \\
            & \ CTC-CMLM &  & 17.2\phantom{0} & \phantom{0}4.49$\times$ & \phantom{0}8.59$\times$ \\
            & \ {\proposed}-CMLM &  & \bf{17.6}\phantom{0} & \phantom{0}4.12$\times$ & \phantom{0}4.61$\times$ \\
            & \ \ + MMT + NAR MT &  & 17.4\phantom{0} & \phantom{0}4.12$\times$ & \phantom{0}4.61$\times$ \\
            \midrule
            
        Cascade AR
            & Cfm ASR $\to$ Trf MT & $2\ymax$ & 17.3\phantom{0} & \phantom{0}0.42$\times$ & \phantom{0}0.79$\times$ \\
        \bottomrule
    \end{tabular}
    }
    \endgroup
\end{table}

For Conformer-based models, we observed large improvements across all models.
The AR model achieved the state-of-the-art BLEU score of 19.2.
The other trend was consistent with that on Must-C.
Conformer-CTC improved upon the Transformer-CTC by 4.1 BLEU, and {\proposed}-CTC further boosted it by 1.0 BLEU.
{\proposed}-CMLM outperformed the CMLM by 1.4 BLEU for $\totaliter=4$ and 1.8 BLEU for $\totaliter=10$, while the enhanced training did not improve it further.

\begin{table*}[t]
    \centering
    \begingroup
    \caption{BLEU scores with \textbf{Transformer encoder} on \underline{Fisher-CallHome Spanish Es$\to$En}. $\ndecar$ was set to 3. Decoding speed was measured on Fisher-\texttt{test} set.}
    \label{tab:taslp2021_results_fisher_transformer}
    \vspace{-1mm}
    \scalebox{1.0}{
    \begin{tabular}{@{}clcccccccc@{}}\toprule
      \multicolumn{2}{c}{\multirow{3}{*}{Model}} & \multirow{3}{*}{$\totaliter$} & \multicolumn{5}{c}{BLEU ($\uparrow$)}& \multicolumn{2}{c}{Speedup ($\uparrow$)} \\ \cmidrule(lr){4-8} \cmidrule(lr){9-10}
      \multicolumn{2}{c}{} & & \multicolumn{3}{c}{Fisher} & \multicolumn{2}{c}{CallHome} & \multirow{2}{*}{GPU} & \multirow{2}{*}{CPU} \\ \cmidrule(lr){4-6} \cmidrule(lr){7-8}
      \multicolumn{2}{c}{} & & dev & dev2 & test & devtest & evltest &  & \\ \midrule
            \multirow{4}{*}{\shortstack{E2E AR}} & \ Transformer ($\beamst=1$) & \multirow{4}{*}{$\ymax$} & 47.3 & 48.5 &	47.5 & 18.0 & 17.3 & \phantom{0}1.84$\times$ & \phantom{0}3.07$\times$ \\
            & \ Transformer ($\beamst=4$) &  & 49.4 & 50.6 & 49.4 & 18.6 & 18.4 & \phantom{0}1.00$\times$ & \phantom{0}1.00$\times$ \\
            & \ Transformer + SeqKD ($\beamst=1$) &  & 49.6 & 49.9 & 49.4 & 18.5 & 18.6 & \phantom{0}1.84$\times$ & \phantom{0}3.07$\times$ \\
            & \ Transformer + SeqKD ($\beamst=4$) &  & \bf{51.1} & \bf{51.4} & \bf{50.8} & \bf{19.6} & \bf{19.2} & \phantom{0}1.00$\times$ & \phantom{0}1.00$\times$ \\
            \midrule
            
            \multirow{10}{*}{\shortstack{E2E NAR}} & \ CTC ($\lengthbeam=1$) & 1 & 45.8 & 46.4 & 46.2 & 15.6 & 16.0 & 22.11$\times$ & 11.55$\times$ \\
            \cdashlinelr{2-10}
            
            & \ CMLM & \multirow{4}{*}{4} & 45.3 & 46.2 & 45.4 & 17.5 & 16.8 & \phantom{0}5.76$\times$ & \phantom{0}4.50$\times$ \\
            & \ SMART &  & 45.3 & 46.1 & 45.9 & 17.2 & 17.0 & \phantom{0}5.76$\times$ & \phantom{0}4.50$\times$ \\
            & \ {\proposed}-CMLM &  & 47.4 & 48.2 & 47.4 & 18.6 & 17.8 & \phantom{0}4.95$\times$ & \phantom{0}4.05$\times$ \\
            & \ \ + MMT + NAR MT &  & \bf{49.1} & \bf{49.9} & \bf{48.5} & \bf{19.0} & \bf{18.9} & \phantom{0}4.95$\times$ & \phantom{0}4.05$\times$ \\
            \cdashlinelr{2-10}
            
            & \ CMLM & \multirow{4}{*}{10} & 47.7 & 48.5 & 48.0 & 18.5 & 18.6 & \phantom{0}3.19$\times$ & \phantom{0}2.47$\times$ \\
            & \ SMART &  & 46.3 & 47.1 & 47.0 & 17.9 & 17.8 & \phantom{0}3.19$\times$ & \phantom{0}2.47$\times$ \\
            & \ {\proposed}-CMLM &  & 49.9 & 50.4 & 49.6 & 19.3 & 18.9 & \phantom{0}2.93$\times$ & \phantom{0}2.34$\times$ \\
            & \ \ + MMT + NAR MT &  & \bf{50.8} & \bf{51.3} & \bf{50.1} & \bf{19.5} & \bf{19.6} & \phantom{0}2.93$\times$ & \phantom{0}2.34$\times$ \\
            \midrule
            
        Cascade AR & Trf ASR $\to$ Trf MT & $2\ymax$ & 41.4 & 43.3 & 42.3 & 	20.4 & 19.9 & \phantom{0}0.50$\times$ & \phantom{0}0.64$\times$ \\
        
        \bottomrule
    \end{tabular}
    }
    \endgroup
\end{table*}

\begin{table*}[t]
    \centering
    \begingroup
    \caption{BLEU scores with \textbf{Conformer encoder} on \underline{Fisher-CallHome Spanish Es$\to$En}}
    \label{tab:taslp2021_results_fisher_conformer}
    \vspace{-1mm}
    \scalebox{1.0}{
    \begin{tabular}{@{}clcccccccc@{}}\toprule
      \multicolumn{2}{c}{\multirow{3}{*}{Model}} & \multirow{3}{*}{$\totaliter$} & \multicolumn{5}{c}{BLEU ($\uparrow$)}& \multicolumn{2}{c}{Speedup ($\uparrow$)} \\ \cmidrule(lr){4-8} \cmidrule(lr){9-10}
      \multicolumn{2}{c}{} & & \multicolumn{3}{c}{Fisher} & \multicolumn{2}{c}{CallHome} & \multirow{2}{*}{GPU} & \multirow{2}{*}{CPU} \\ \cmidrule(lr){4-6} \cmidrule(lr){7-8}
      \multicolumn{2}{c}{} & & dev & dev2 & test & devtest & evltest &  & \\ \midrule
            \multirow{4}{*}{\shortstack{E2E AR}} & \ Conformer ($\beamst=1$) &  \multirow{4}{*}{$\ymax$} & 53.3 & 53.9 & 52.5 & 20.6 & 20.8 & \phantom{0}1.82$\times$ & \phantom{0}2.71$\times$ \\
            & \ Conformer ($\beamst=4$) &  & 54.4 & 55.1 & 53.6 & 21.1 & 21.1 & \phantom{0}1.00$\times$ & \phantom{0}1.00$\times$ \\
            & \ Conformer + SeqKD ($\beamst=1$) &  & 54.1 & 54.6 & 53.9 & 21.4 & \bf{21.5} & \phantom{0}1.82$\times$ & \phantom{0}2.71$\times$ \\
            & \ Conformer + SeqKD ($\beamst=4$) &  & \bf{54.7} & \bf{55.4} & \bf{54.1} & \bf{21.5} & 21.0 & \phantom{0}1.00$\times$ & \phantom{0}1.00$\times$ \\
            \midrule
            
            \multirow{10}{*}{\shortstack{E2E NAR}} & \ CTC ($\lengthbeam=1$) & \multirow{2}{*}{1} & 51.0 & 51.6 & 50.8 & 18.0 & 18.7 & 11.80$\times$ & \phantom{0}6.91$\times$ \\
        
            & \ {\proposed}-CTC ($\lengthbeam=20$) &  & \bf{54.0} & \bf{54.8} & \bf{54.1} & \bf{21.0} & \bf{20.8} & \phantom{0}1.09$\times$ & \phantom{0}2.80$\times$ \\  %
           \cdashlinelr{2-10}

            & \ CMLM & \multirow{4}{*}{4} & 47.7 & 48.1 & 46.9 & 18.8 & 18.5 & \phantom{0}4.70$\times$ & \phantom{0}3.68$\times$ \\
            & \ CTC-CMLM &  & 50.9 & 51.4 & 50.5 & 17.7 & 17.9 & \phantom{0}4.79$\times$ & \phantom{0}4.96$\times$ \\
            & \ {\proposed}-CMLM &  & \bf{50.3} & 50.7 & 48.9 & 20.0 & 18.9 & \phantom{0}4.18$\times$ & \phantom{0}3.44$\times$ \\
            & \ \ + MMT + NAR MT &  &\bf{50.3} & \bf{50.7} & \bf{49.0} & \bf{20.3} & \bf{20.0} & \phantom{0}4.18$\times$ & \phantom{0}3.44$\times$ \\
            \cdashlinelr{2-10}

            & \ CMLM & \multirow{4}{*}{10} & 49.8 & 49.7 & 48.9 & 19.7 & 19.4 & \phantom{0}2.89$\times$ & \phantom{0}2.30$\times$ \\
            & \ CTC-CMLM &  & 51.3 & 51.7 & 50.7 & 17.9 & 18.3 & \phantom{0}3.40$\times$ & \phantom{0}3.89$\times$ \\
            & \ {\proposed}-CMLM &  & \bf{51.8} & \bf{52.3} & 50.9 & 20.7 & 20.0 & \phantom{0}2.70$\times$ & \phantom{0}2.20$\times$ \\
            & \ \ + MMT + NAR MT &  & 51.3 & 52.2 & \bf{51.2} & \bf{20.9} & \bf{20.4} & \phantom{0}2.70$\times$ & \phantom{0}2.20$\times$ \\
            \midrule
            
        Cascade AR & Cfm ASR $\to$ Trf MT & $2\ymax$ & 42.6 & 44.4 & 43.2 & 21.7 & 21.0 & \phantom{0}0.50$\times$ & \phantom{0}0.66$\times$ \\
        \bottomrule
    \end{tabular}
    }
    \endgroup
\end{table*}

\subsection{Fisher-CallHome Spanish}
The results with the Transformer and Conformer encoders on Fisher-CallHome Spanish are listed in Tables~\ref{tab:taslp2021_results_fisher_transformer} and ~\ref{tab:taslp2021_results_fisher_conformer}, respectively.
Note that the CallHome sets are out-of-domain because the models were trained in the Fisher domain.
We confirmed similar trends for Must-C and Libri-trans.
However, unlike previous experiments, we observed that Conformer-based {\proposed}-CMLM outperformed Conformer-CTC on the CallHome test sets by a large margin.
Considering the fact that the word error rates on the CallHome test sets were much larger than those on the Fisher test sets (40\% vs. 20\%~\cite{karita2019comparative,guo2021recent}), we can conclude that {\proposed} is more robust in noisy acoustic conditions.

\section{Analysis}

\begin{table}[t]
    \centering
    \caption{Effective number of auxiliary AR decoders $\ndecar$ for {\proposed}-CMLM on \texttt{dev} sets of Must-C En$\to$De, Libri-trans, and Fisher-CallHome Spanish. \textbf{Conformer encoder} was used. All models used parallel AR rescoring. No enhanced training method, such as MMT, was used.}
    \label{tab:taslp2021_results_number_ar_decoder}
    \vspace{-1mm}
    \begingroup
    \scalebox{1.0}{
    \begin{tabular}{@{}ccccccc@{}}\toprule
    \multirow{3}{*}{$\ndecar$} & \multicolumn{6}{c}{BLEU ($\uparrow$)} \\  \cmidrule(lr){2-7}
    & \multicolumn{2}{c}{Must-C} & \multicolumn{2}{c}{Libri-trans} & \multicolumn{2}{c}{Fisher} \\ \cmidrule(lr){2-3} \cmidrule(lr){4-5} \cmidrule(lr){6-7}
    & $\totaliter=4$ & $\totaliter=10$ & $\totaliter=4$ & $\totaliter=10$ & $\totaliter=4$ & $\totaliter=10$ \\
    \midrule
       0 & 21.4 & 22.7 & 16.2 & 16.9 & 47.7 & 49.8 \\ 
       \cmidrule(lr){1-7}
       1 & \bf{22.2} & \bf{23.2} & \bf{17.4} & \bf{18.4} & 48.7 & 50.1 \\
       2 & 21.9 & 23.1 & 17.3 & 18.2 & 49.2 & 51.0 \\
       3 & 22.0 & \bf{23.2} & 16.8 & 18.1 & \bf{50.3} & \bf{51.8} \\
       4 & 21.0 & 22.4 & \bf{17.6} & \bf{18.5} & 49.5 & 51.6 \\
       5 & 21.8 & 23.1 & 17.3 & 18.4 & 47.9 & 49.5 \\
       6 & 21.9 & 22.8 & 16.9 & 17.8 & 49.1 & 51.1 \\
     \bottomrule
    \end{tabular}
    }
    \endgroup
\end{table}

\subsection{Effective number of auxiliary AR decoder layers}
We first investigated the effective number of auxiliary AR decoder layers $\ndecar$ for {\proposed}-CMLM.
We increased $\ndecar$ from zero to six.\footnote{In our previous study~\cite{inaguma2021orthros}, we used $\ndecar=6$ and did not study its impact on BLEU scores.}
We used the Conformer encoder without any enhanced training method.
The results on the \texttt{dev} set of Must-C En$\to$De, Libri-trans, and Fisher-CallHome Spanish in Table~\ref{tab:taslp2021_results_number_ar_decoder} indicate that a shallow AR decoder works well across corpora.
Specifically, $\ndecar=1$ was best on Must-C, $\ndecar=4$ was best on Libri-trans while $\ndecar=1$ showed comparable BLEU scores.
However, Fisher-CallHome Spanish reached a peak at $\ndecar=3$.
This was probably because the input speech was much noisy compared with other corpora; therefore, stacking multiple AR decoder layers was more effective for providing better sequence-level scores.
Otherwise, a single layer was enough, which minimized the additional decoding cost by the rescoring.
Hence, we used $\ndecar=3$ on Fisher-CallHome Spanish and $\ndecar=1$ on the other corpora in the other experiments, regardless of the encoder type.
We also used the same $\ndecar$ for {\proposed}-CTC.

\begin{table*}[t]
    \centering
    \caption{Ablation study of parallel AR rescoring and enhanced training methods on \texttt{dev} sets of Must-C En$\to$De, Libri-trans, and Fisher-CallHome Spanish. \textbf{Transformer encoder} was used.}
    \label{tab:taslp2021_results_ablation_study}
    \vspace{-1mm}
    \begingroup
    \scalebox{1.0}{
        \begin{tabular}{@{}clcccccc@{}}\toprule
        \multirow{3}{*}{ID} & \multirow{3}{*}{Model} & \multicolumn{6}{c}{BLEU ($\uparrow$)} \\
         \cmidrule(lr){3-8}
         & & \multicolumn{2}{c}{Must-C} & \multicolumn{2}{c}{Libri-trans} & \multicolumn{2}{c}{Fisher}  \\
         \cmidrule(lr){3-4} \cmidrule(lr){5-6} \cmidrule(lr){7-8}
         & & $\totaliter=4$ & $\totaliter=10$ & $\totaliter=4$ & $\totaliter=10$ & $\totaliter=4$ & $\totaliter=10$ \\
         \midrule
          \texttt{A1} & CMLM & 19.8 & 21.4 & 14.6 & 15.8 & 45.3 & 47.7 \\
          \cmidrule{1-8}
          
          \texttt{A2} & {\proposed}-CMLM & 21.1 & 22.0 & 15.7 & 16.9 & 47.4 & 49.9 \\
          \texttt{A3} & \ - parallel AR rescoring & 20.4 & 21.3 & 15.1 & 16.3 & 45.7 & 48.3 \\
          \cmidrule{1-8}
           
          \texttt{A4} & {\proposed}-CMLM + MMT + NAR MT & \bf{21.9} & 22.8 & 16.3 & \bf{17.3} & \bf{49.1} & \bf{50.8} \\
          \texttt{A5} & \ - parallel AR rescoring & 21.5 & 22.4 & 15.7 & 16.7 & 47.2 & 49.0 \\
          \texttt{A6} & \ - MMT ($\nmask=2$) & 21.8 & \bf{23.0} & 15.6 & 16.3 & 48.6 & 50.2 \\
          \texttt{A7} & \ - NAR MT & 21.2 & 22.4 & \bf{16.4} & 17.2 & 47.5 & 49.9 \\
          \texttt{A8} & \ \ \ + MMT ($\nmask=3$) & 21.8 & 22.6 & 16.1 & 17.2 & 47.9 & 50.1 \\
          \texttt{A9} & \ \ \ + MMT ($\nmask=4$) & 21.4 & 22.4 & 16.1 & 17.0 & 48.7 & 50.4 \\

          \bottomrule
        \end{tabular}
    }
    \endgroup
\end{table*}

\subsection{Ablation study}\label{ssec:ablation_study_orthros}
We conducted an ablation study of {\proposed}-CMLM in terms of parallel AR rescoring and the enhanced training methods on the \texttt{dev} sets of Must-C En$\to$De, Libri-trans, and Fisher-CallHome Spanish shown in Table~\ref{tab:taslp2021_results_ablation_study}.
We used the Transformer encoder and first observed that joint training with an auxiliary shallow AR decoder improved the BLEU scores (\texttt{A1} vs. \texttt{A3}), except for $\totaliter=10$ on Must-C.
\texttt{A3} was trained jointly with the auxiliary AR decoder but did not use it during inference.
Parallel AR rescoring further improved the BLEU scores in all settings (\texttt{A2} vs. \texttt{A3}).

The effectiveness of parallel AR rescoring remained even when using the enhanced training combining MMT ($\nmask=2$) and the auxiliary NAR-MT task (\texttt{A4} vs. \texttt{A5}).
Each of these training methods was beneficial (\texttt{A4} vs. \texttt{A6} vs. \texttt{A7}).
Because \texttt{A4} increased the number of masks for the CMLM training by a factor of 4 per sample, we also investigated increasing $\nmask$ in MMT without the auxiliary NAR-MT task (\texttt{A9}).
However, we confirmed that this was less effective, indicating that providing source information from different modalities was helpful to some extent.

\begin{table}[t]
    \centering
    \tabcolsep 2pt
    \begingroup
    \caption{BLEU scores of \textbf{large} models with $(\dmodel, \nhead)=(512,8)$ on \underline{Must-C En$\to$De} \texttt{tst-COMMON}. \textbf{Transformer encoder} was used.}
    \label{tab:taslp2021_results_mustc_large}
    \vspace{-1mm}
    \scalebox{1.0}{
    \begin{tabular}{@{}clcccc@{}}\toprule
      \multicolumn{2}{c}{\multirow{2}{*}{Model}} & \multirow{2}{*}{$\totaliter$} & \multirow{2}{*}{BLEU ($\uparrow$)} & \multicolumn{2}{c}{Speedup ($\uparrow$)} \\ \cmidrule(lr){5-6}
      \multicolumn{2}{c}{} &  & & GPU & CPU \\ \midrule
            & Transformer AR + SeqKD ($\beamst=1$) & \multirow{2}{*}{$\ymax$} & 24.0 & 1.62$\times$ & 3.30$\times$ \\
            & Transformer AR + SeqKD ($\beamst=4$) &  & 24.7 & 1.00$\times$ & 1.00$\times$ \\ \cmidrule(lr){1-6}
            
            & {\proposed}-CMLM & \multirow{2}{*}{4} & 23.0 & \multirow{2}{*}{5.89$\times$} & \multirow{2}{*}{5.63$\times$} \\
            & \ + MMT + NAR MT &  & \bf{23.8} \\
            \cmidrule(lr){1-6}

            & {\proposed}-CMLM & \multirow{2}{*}{10} & 24.0 & \multirow{2}{*}{3.02$\times$} & \multirow{2}{*}{2.92$\times$} \\
            & \ + MMT + NAR MT &  & \bf{24.4} \\
        \bottomrule
    \end{tabular}
    }
    \endgroup
\end{table}

\subsection{Large model}
In the above experiments, we used $(\dmodel, \nhead)=(256, 4)$ in all models.
They are much smaller than those used in the NAR MT literature~\cite{gu-kong-2021-fully}, where $(\dmodel, \nhead)=(512, 8)$ was typically used.
Therefore, our baseline AR model was relatively lightweight.
This was because using a large AR model did not lead to a notable BLEU improvement in our preliminary experiments due to the limited parallel data in the ST corpora.
However, when additional data are available, increasing model capacity would be typically beneficial.
Therefore, we also compared decoding speeds of AR and NAR models based on a large Transformer architecture.
We used $(\dmodel, \nhead)=(512, 8)$ in both the encoder and decoder architectures of AR and NAR models, except that the size of the auxiliary AR decoder in {\proposed}-CMLM was kept to $(\dmodel, \nhead)=(256, 4)$.
The results on Must-C En$\to$De \texttt{tst-COMMON} in Table~\ref{tab:taslp2021_results_mustc_large} indicate that using the large architecture improved the BLUE score of the AR model by only 0.4.
The inference speed was not so different from that in Table~\ref{tab:taslp2021_results_mustc_transformer}, but we observed a large BLEU improvement in {\proposed}-CMLM by increasing the model capacity.
Compared with Table~\ref{tab:taslp2021_results_mustc_transformer}, it improved BLEU scores by 1.2 and 1.1 when $\totaliter=4$ and $\totaliter=10$, respectively.
The quality was further boosted by the enhanced training by 0.8 and 0.4 for $\totaliter=4$ and $\totaliter=10$, respectively.
Finally, the enhanced {\proposed}-CMLM almost matched the strong AR model trained with SeqKD in quality with 3$\times$ faster decoding.

\begin{figure}[t]
    \centering
    \includegraphics[width=1.0\linewidth]{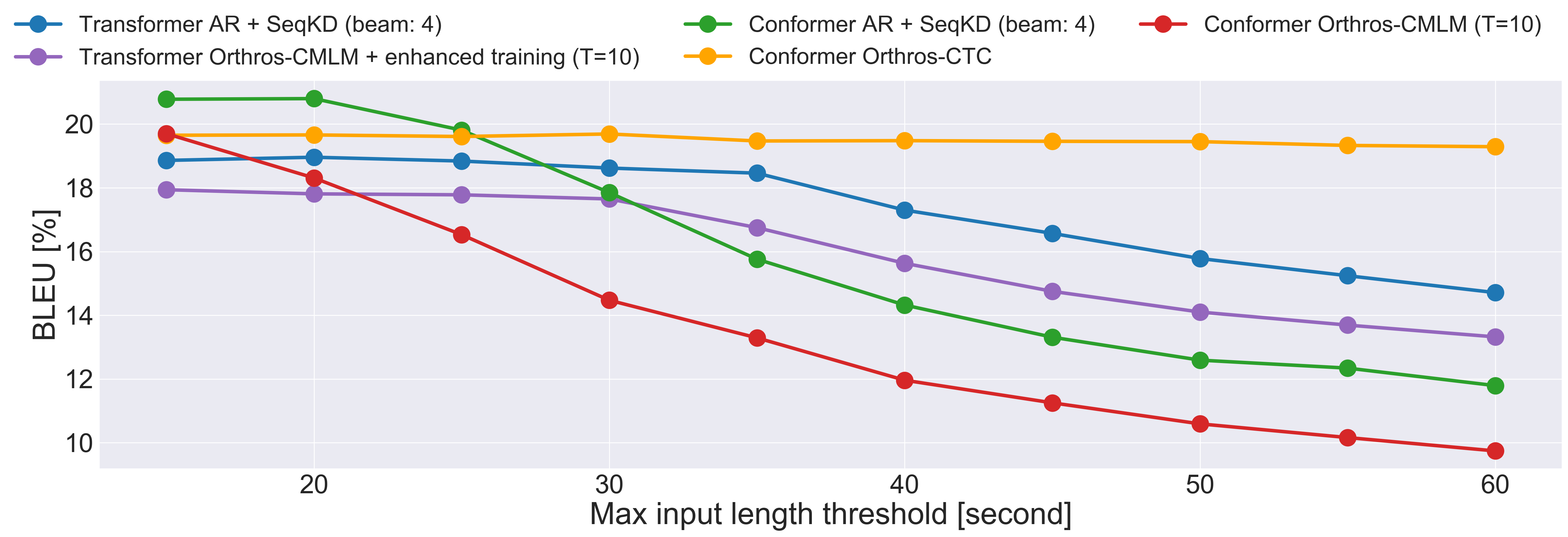}
    \caption{Robustness against long-form speech on IWSLT \texttt{tst2019} set. X-axis denotes maximum input length threshold in segment-merging algorithm~\cite{inaguma-etal-2021-espnet}.}
    \label{fig:long_form}
\end{figure}

\subsection{Robustness against long-form speech}
In speech translation, audio segmentation has a large impact on the final translation quality~\cite{gaido2020contextualized,pham2020relative,potapczyk-przybysz-2020-srpols,gaido2021beyond,inaguma-etal-2021-espnet} because acoustic-driven segmentation does not necessarily correspond to sentence segmentation based on punctuation marks.
Therefore, it is important to translate long-form speech robustly by incorporating long context~\cite{inaguma-etal-2021-espnet}.
Therefore, we investigated the robustness of E2E-ST models against long-form data on the International Conference on Spoken Language Translation (IWSLT) \texttt{tst2019} set (En$\to$De).\footnote{The IWSLT test sets have reference translation over the entire session, unlike Must-C.}
Following the segment-merging algorithm~\cite{inaguma-etal-2021-espnet}, we split the entire speech using a neural voice activity detection model~\cite{bredin2020pyannote} then concatenated multiple adjacent segments until reaching the desired input length.\footnote{We did not merge adjacent segments with interval above 100ms.}
Figure~\ref{fig:long_form} shows BLEU scores as a function of the maximum input length threshold for the segment merging.
We evaluated models trained on Must-C En$\to$De.
Note that we did not change any of the decoding hyperparameters of the E2E-ST models for this experiment.
We observed that both the AR model and {\proposed}-CMLM were not robust against long-form speech, regardless of the encoder architecture.
In contrast, {\proposed}-CTC could robustly translate long-form speech up to 60 s, outperforming the Conformer AR model on speech longer than 30 s.

\begin{figure}[t]
    \centering
    \includegraphics[width=1.0\linewidth]{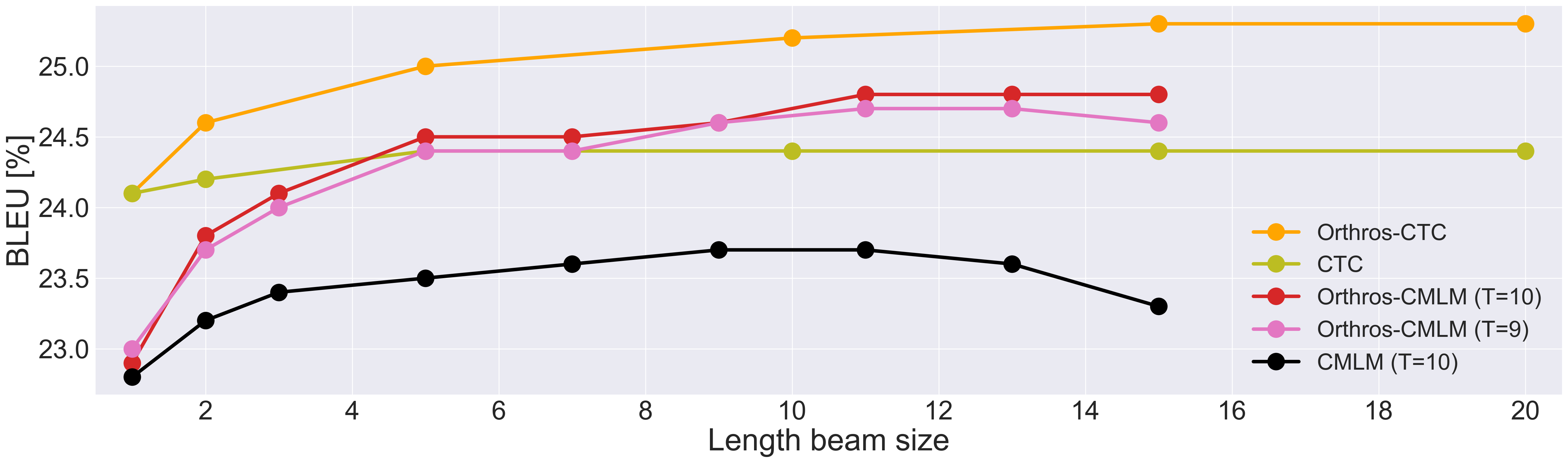}
    \caption{Effectiveness of length beam size $\lengthbeam$ with $\totaliter=10$ on Must-C En$\to$De \texttt{tst-COMMON}. \textbf{Conformer encoder} was used.}
    \label{fig:taslp2021_length_beam}
\end{figure}

\subsection{Searchability}
In this section, we analyze the searchability of {\proposed} by changing a length beam size $\lengthbeam$.
We investigate how it impacts the quality of hypotheses.

\vspace{2mm}
\subsubsection{Length beam size}
Figure~\ref{fig:taslp2021_length_beam} shows BLEU scores as a function of length beam size $\lengthbeam$ with $\totaliter=10$ on Must-C En$\to$De \texttt{tst-COMMON}.
We used the Conformer encoder and confirmed that parallel AR rescoring consistently improved the quality of NAR models when $\lengthbeam > 1$ and the gain increased using a larger $\lengthbeam$.
The baseline CMLM and CTC-based model did not improve by increasing $\lengthbeam$ because the NAR decoders could not provide effective sequence-level scores.
Because parallel AR rescoring introduces an additional iteration in terms of $\totaliter$, we also investigated reducing $\totaliter$ in {\proposed}-CMLM by 1.
We confirmed that BLEU scores of {\proposed}-CMLM with $\totaliter=9$ was much better than those of CMLM with $\totaliter$ regardless of $\lengthbeam$ and the gap was enlarged as $\lengthbeam$ increased.
Therefore, we conclude that parallel AR rescoring improves the searchability of NAR models.

\begin{table}[t]
    \centering
    \begingroup
    \caption{Oracle BLEU scores on \underline{Must-C En$\to$De} \texttt{tst-COMMON}. Numbers inside brackets denote gains from parallel AR rescoring.}
    \label{tab:taslp2021_results_oracle}
    \vspace{-1mm}
    \scalebox{1.0}{
    \begin{tabular}{@{}lcccc@{}}\toprule
      Model & $\totaliter$ & $\lengthbeam$ & \textbf{Oracle BLEU} ($\bm{\Delta}$) ($\uparrow$) \\ \midrule
            \textbf{Transformer encoder} \\
            \ {\proposed}-CMLM & 4 & 5 & 24.6 (+2.8) \\
            & 4 & 9 & 26.1 (+3.9) \\
            & 10 & 5 & 25.9 (+3.0) \\
            & 10 & 9 & 27.2 (+4.2) \\
            \cdashlinelr{1-5}
            \ \ + MMT + NAR MT & 4 & 5 & 25.7 (+3.2) \\
            & 4 & 9 & 27.2 (+4.4) \\
            & 10 & 5 & 26.9 (+3.4) \\
            & 10 & 9 & 28.4 (+4.6) \\
            \cmidrule{1-5}
            \ \ + large & 4 & 5 & 26.1 (+3.1) \\
            & 4 & 9 & 27.5 (+4.4) \\ %
            & 10 & 5 & 27.4 (+3.4) \\
            & 10 & 9 & 28.8 (+4.6) \\
            \midrule
            
            \textbf{Conformer encoder} \\
            \ {\proposed}-CTC & 1 & 20 & 28.6 (+3.3) \\
            \cdashlinelr{1-5}
            \ {\proposed}-CMLM & 4 & 5 & 26.2 (+2.4) \\
            & 4 & 9 & 27.5 (+4.1) \\ 
            & 10 & 5 & 27.1 (+2.5) \\
            & 10 & 9 & 28.5 (+4.0) \\
        \bottomrule
    \end{tabular}
    }
    \endgroup
\end{table}

\vspace{2mm}
\subsubsection{Oracle BLEU}
We next investigated oracle BLEU scores, where the best hypothesis was selected from candidates generated from the NAR decoder based on the sentence-level BLEU score with the reference translation.
It is regarded as the upper bound that can be achieved by parallel AR rescoring.
 We conducted this experiment to identify a room for improvement of translation quality of NAR models.
Table~\ref{tab:taslp2021_results_oracle} indicates that increasing the number of length candidates can generate translations of higher quality than refining the predictions for more iterations.
However, NAR models were still affected by selecting a better translation even with parallel AR rescoring.
This was more severe when using a large $\lengthbeam$.
Better training and architecture also led to higher oracle BLEU scores.

\section{Conclusions}\label{sec:conclution}
To increase the decoding speed of E2E-ST models, we proposed {\proposed}, which introduces an auxiliary shallow AR decoder on top of the shared encoder to assist the NAR decoder and optimizes the entire network end-to-end.
The AR decoder is used for rescoring outputs from the NAR decoder with a very small additional decoding cost.
We investigated the CMLM and CTC as the NAR decoder for {\proposed}.
We also compared Transformer and Conformer encoder architectures.
We introduced MMT and joint training with an auxiliary text-input NAR-MT task to enhance the CMLM decoder.
Experimental evaluations on three benchmark corpora including six language pairs confirmed the consistent effectiveness of {\proposed}; parallel AR rescoring improved the BLUE scores of the NAR model regardless of the encoder and decoder topologies.
Enhanced training methods improved the quality of {\proposed}-CMLM by a large margin when the Transformer encoder was used.
When increasing the model capacity, the enhanced {\proposed}-CMLM matched the strong large AR model trained with SeqKD in terms of BLEU score with 3$\times$ faster decoding.
The Conformer encoder improved the overall quality across models while the effectiveness of the parallel AR rescoring was maintained.
{\proposed}-CTC showed the best BLEU score while achieving a 3.63$\times$ increase in decoding speed on a CPU compared with the AR model.

For future work, we will further improve the translation quality of Conformer-CTC by adopting a deep encoder architecture~\cite{kasai2021deep}.
We believe that improving the selection of a better hypothesis in NAR models would bridge the quality gap between AR and NAR models.

\bibliographystyle{IEEEtran}
\bibliography{reference}

\ifCLASSOPTIONcaptionsoff
  \newpage
\fi

\end{document}